\documentclass[12pt]{spieman}  
\usepackage{amsmath,amsfonts,amssymb,aas_macros}
\usepackage{graphicx}
\usepackage{setspace}
\usepackage{tocloft}

\title{Improving the Thermal Stability of a CCD Through Clocking}

\author[a,*]{Cullen H. Blake}
\author[b]{Dan Li}
\author[c]{Joseph R. Tufts}
\author[d]{Joe Ninan}
\author[c]{Suvrath Mahadevan}
\author[d]{Chad Bender}
\author[c]{Fred R. Hearty}
\author[c]{Andy Monson}
\author[a]{Mark Giovinazzi}
\affil[a]{University of Pennsylvania, Department of Physics and Astronomy, Philadelphia, PA 19104}
\affil[b]{National Optical Astronomical Observatory, Tucson, AZ, 85719}
\affil[c]{STA, Inc, San Clemente, CA 92673}
\affil[c]{Penn State University, Department of Astronomy, State College, PA 16802}
\affil[d]{Steward Observatory, Tucson, AZ, 85719}

\cftpagenumbersoff{figure}
\cftpagenumbersoff{table} 
\begin{document} 
\maketitle

\begin{abstract}

Modern precise radial velocity spectrometers are designed to infer the existence of planets orbiting other stars by measuring few-nm shifts in the positions of stellar spectral lines recorded at high spectral resolution on a large-area digital detector. While the spectrometer may be highly stabilized in terms of temperature, the detector itself may undergo changes in temperature during readout that are an order of magnitude or more larger than the other opto-mechanical components within the instrument. These variations in detector temperature can translate directly into systematic measurement errors. We explore a technique for reducing the amplitude of CCD temperature variations by shuffling charge within a pixel in the parallel direction during integration. We find that this ``dither clocking" mode greatly reduces temperature variations in the CCDs being tested for the NEID spectrometer. We investigate several potential negative effects this clocking scheme could have on the underlying spectral data.  

\end{abstract}


{\noindent \footnotesize\textbf{*}Cullen H. Blake,  \linkable{chblake@sas.upenn.edu} }

\begin{spacing}{2}   

\section{Introduction}
\label{sect:intro}  
The next generation of astronomical spectrometers designed for detecting Earth-mass exoplanets seek to measure the radial velocities (RVs) of stars with a precision of 10-50 cm s$^{-1}$. For a typical instrument design, this corresponds to measuring an ensemble Doppler shift in the wavelengths of stellar spectral lines on the focal plane of approximately 0.001~nm. At this level of measurement precision, a host of errors, both instrumental and astrophysical, become important\cite{Halverson2016}. To minimize some of these sources of instrumental measurement error, it is important to keep the opto-mechanical components of the spectrometer as stable as possible. This usually means operating in high vacuum, and controlling the temperature of the entire optical system to the mK level or better\cite{Robertson2016}. The detectors themselves, either CCDs in the optical or HgCdTe arrays in the near infrared, may represent the largest sources of thermal variations in the entire instrument. This stems from the fact that the detectors and vacuum electronics may consume different amounts of power during idle, integration, and readout. 

The impact of the detector's thermal variations on the rest of the instrument may be important, for example through radiative coupling of the detector to adjacent optics. At the same time, it is also possible that temperature changes in the detector package lead directly to physical deformations of the array, which translate directly into measured RV shifts that will need to be calibrated out. While it is challenging to directly simulate these effects for a specific detector package and mounting scheme, given the typical dispersion properties of a high-resolution spectrometer, a physical shift of the focal plane array of just 2~nm represents a systematic error larger than the Doppler shift of the stellar spectral lines resulting from an orbiting Earth-mass planet in a one year orbit. 

Mounting the detector so that it is cooled efficiently, and therefore dissipates heat efficiently, is very important. But, as large-format detectors approach almost 10 cm by 10 cm in size, the thermal time constant of the device itself necessarily becomes long. At the same time, it becomes a challenge to maintain the temperature uniformity of the large detector package given the wide range of integration times typical for an instrument in operation at an astronomical observatory. Given the large number of output channels, the package could itself support temperature inhomogeneities that change with time. Actively controlling the temperature of the CCD package directly, with heaters affixed to the package, may be impractical for a number of reasons. We explore clocking techniques designed to minimize the amplitude of any thermal variations of a four-phase CCD under normal operating conditions and investigate the impact these clocking modes might have on the resulting high-resolution stellar spectra and our ability to infer precise RV measurements from those spectra.   

\section{The NEID Spectrometer Detector System}

NEID\footnote{NEID is a word that means ``to see" or ``to discover/visualize" in the language of the Tohono O'odham, on whose land Kitt Peak National Observatory sits} is a high-resolution, optical (380 to 930 nm) Doppler spectrometer being built for the 3.5-m Wisconsin-Indiana-Yale-NOAO (WIYN) telescope at Kitt Peak National Observatory. The spectrometer is a cross-dispersed, white-pupil design and achieves spectral resolution of R=120,000 while simultaneously recording more than 95 spectral orders on a large-format CCD detector\cite{Schwab2016}. The instrument is contained in a vacuum vessel that maintains pressure below 10$^{-6}$~Torr with an operating temperature of 300~K, which is actively maintained to better than $\pm1$~mK by a temperature control system\cite{Robertson2016}. The CCD detector, which is operated at 173~K, is contained in this same vacuum envelope. The detector is a large-format device from Teledyne/e2v containing 9216x9232 pixels with a 10~$\mu$m pitch. In order to optimize the red response of the system, and to reduce fringing effects at those wavelengths, NEID employs the deep depletion variant of this CCD290-99 device (40$\mu$m epi layer) with the Astro Multi-2 AR coating. The device has 16 independent output channels, which we read out using an Archon controller\cite{Archon} and custom preamplifier boards designed by STA, Inc. Archon is an FPGA-based CCD readout system that also provides a scripting environment for CCD control and sophisticated temperature control capabilities. Inside the vacuum, we perform AC coupling and JFET buffering at the warm end of CCD flex cables, then convert the signal to a true differential output for transmission of the analog signal across a pair of 49-pair shielded twinax cables, each 1.15~m long. We take advantage of the CCD dummy outputs to reject common-mode artifacts across the 16 output channels. While it is possible to read out the device at speeds up to 1~MHz, we have adopted a standard readout mode of 250 kHz, which is a good balance between read noise and total readout time. 

\begin{figure}
\begin{center}
\begin{tabular}{c}
\includegraphics[height=10cm]{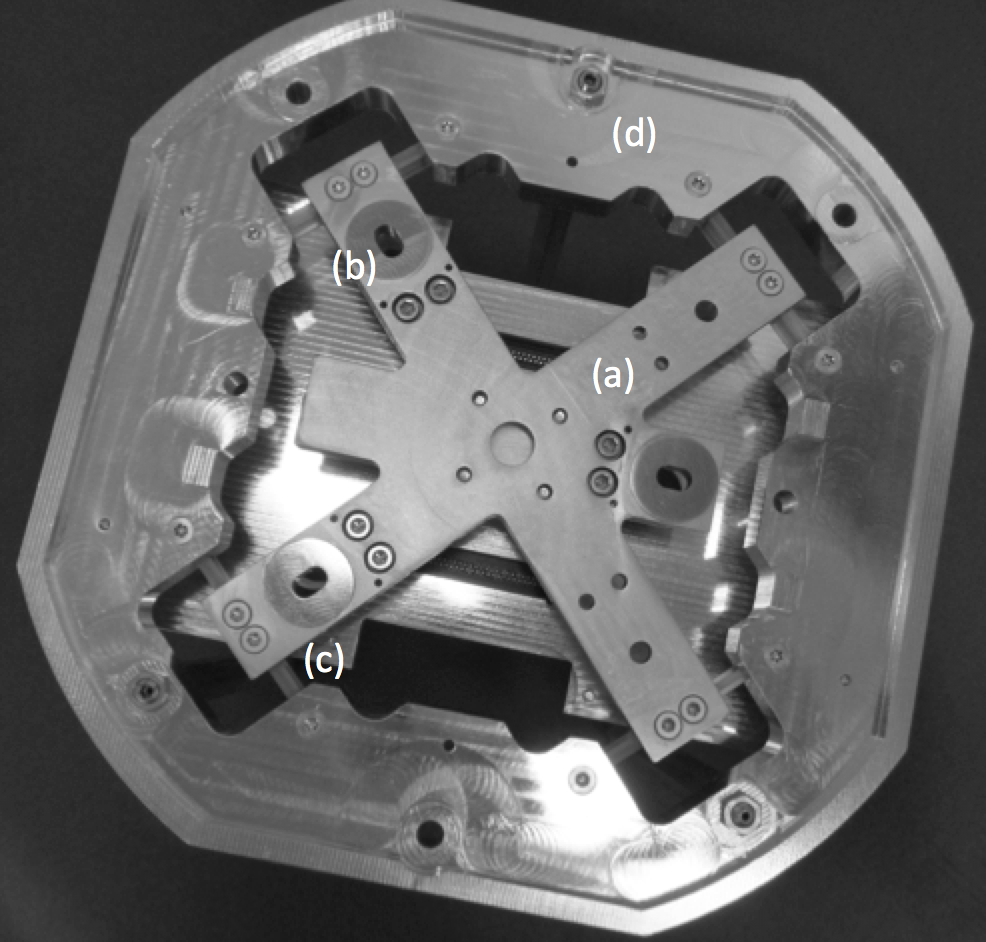}
\end{tabular}
\end{center}
\caption 
{ \label{fig:cold_block}
(a) The Mo cold block to which we mount the CCD290-99. (b) Polished pads on the cold block that mate to the Invar pads on the CCD package. (c) Ultem 2000 spacers that hold the cold block within the Al mounting structure and provide thermal and mechanical isolation. The Al mounting structure (d) attaches directly to the camera lens system within the NEID spectrometer.} 
\end{figure}

The CCD290-99 is contained in a SiC package that has three mounting points for attaching the device to the mounting structure\footnote{https://www.teledyne-e2v.com/shared/content/resources/File/Astronomy/1897.pdf}. Each mounting point consists of a high-precision Invar 36 pad, approximately 4~mm thick and 19~mm in diameter, and an Invar 36 stud which is threaded into the SiC package. The Invar pads are precision polished to provide a mounting plane that is very close to parallel to the active surface of the CCD. We mount the CCD directly on a Mo cold block, shown in Figure \ref{fig:cold_block}, using custom-made, hemispherical M5 nuts that allow the CCD pads to translate slightly due to thermal expansion effects during cool down. Mo was chosen as the material for the CCD mounting block since it has good thermal conductivity and is a good match to the coefficient of thermal expansion of the SiC CCD package. In the CCD test Dewar, the temperature of the Mo cold block is actively controlled by the Archon through an embedded 25~W heater and PT-100 RTD. In NEID, the temperature of the CCD package is actively controlled by Archon using the heater on the Mo cold block. The Mo block is cooled through a tapered copper rod that is attached to the center of the cold block on one end and linked, through a series of copper straps, to the LN2 reservoir of the NEID instrument or test Dewar. Thermal contact to the detector is only through the interface of the Invar pads on the CCD package to the corresponding polished surfaces on the Mo cold block. We monitor the temperature of the CCD package through a PT-100 RTD that is attached, as part of a small Al block, directly to the center of the back of the CCD package using a single M4 screw into a threaded insert that comes installed on the CCD. The CCD package RTD is also read out using Archon, but we have no active heat source on the CCD package.  To improve the cooling properties of the CCD mount system, a thin film of Nye Lubricants NyeTorr 6370EL UHV grease was applied between the Invar pads on the CCD package and the corresponding pads on the Mo block when installed in NEID. A cross section of the three-dimensional model of the NEID CCD mount system is shown in Figure 2.


We carried out an extensive CCD characterization effort using an LN2 cooled IR Labs ND8 test Dewar to house the CCD in essentially the same mounting configuration as in NEID (though no UHV grease was used inside the test Dewar). With less radiation shielding around the CCD and long-term variations in the LN2 fill level of the test Dewar, this CCD testing environment is not as intrinsically stable as NEID. However, this test Dewar enabled very efficient testing of different CCD operational modes, as well as a large number of opto-mechanical tests of the CCD.

\section{Operation of the NIED Detector}

The NEID detector is controlled by Archon using a custom set of scripts to execute timing and voltage commands. We found that the nominal timing parameters described in the CCD290-99 data sheet provided good performance and so we used these parameters to define the nominal clocking pattern. Reading out the device generates a substantial amount of heat, and the device data sheet indicates that up to 100 mW of transient heat may be generated during the CCD readout through the clocking of the parallel and serial registers. There is also a substantial static heat load from the amplifiers estimated at 800 mW, but this does not vary while the CCD is powered on. Given the mass of the CCD package, the thermal properties of SiC, and the approximate time to read out a frame in our 250 kHz mode, we estimated that variable heat loads of this size could lead to changes in the CCD package temperature at the 10~mK level. This is more than a factor of 10 larger than our target absolute CCD temperature stability of $\pm1$~mK, which is set by the design goal of achieving overall intrinsic instrument stability corresponding to 10 cm s$^{-1}$ in RV. Furthermore, the CCD itself is a large cold sink in our warm instrument, and radiative coupling between the spectrometer camera lens system and the CCD, with an area of 85 cm$^{2}$ and undergoing temperature variations at the 10~mK level, was observed to destabilize the entire instrument with a long thermal equilibration timescale. 

We operate the CCD in a standard mode where the parallel and serial registers are constantly being clocked while idle, and the serial registers are also clocked during integration. This means that during integration the heat output of the CCD package decreases as the parallel registers cease to clock. The motivation for this approach was that it should be simpler to control the CCD temperature by adding heat during integration rather than removing heat. Since we are only controlling the temperature of the Mo cold block to which the CCD is attached, the thermal time constant of the CCD package reaction to changes in the cold block may be long compared to integration times. This makes maintaining the thermal stability of the CCD package using this approach a challenge. For example, Figure 3 shows the CCD package temperature within our test Dewar over several hours during a series of calibration lamp exposures of different lengths in standard readout mode. During integration, we see a decrease in CCD package temperature of approximately 20~mK on top of longer timescale changes corresponding to the varying fill level of the LN2 test Dewar.  During this period, the temperature of the Mo cold block, which is controlled directly by Archon in the test Dewar, is stable to better than $\pm1$~mK.


While the amplitude of these CCD temperature variations could be reduced by improving the thermal contact between the CCD package and the Mo cold block, the relatively poor thermal contact between the hard surfaces of the Invar pads on the CCD, the SiC package, and the polished cold block makes this difficult without directly attaching additional cold strapping to the CCD package. The design of the CCD package and our mounting system make this difficult, and increasing the direct physical contact between the CCD package and the cold linkage increases the likelihood that mechanical variations in the system (due to LN2 fill level, for example) may manifest as physical displacements in the focal plane. Instead, we explored ways to reduce the overall variations in the heat output of the CCD by exercising the parallel clocks during integration. 

\section{Dither Clocking}

We present a technique, which we call ``dither clocking", where the voltages of two phases of our four-phase CCD are modulated during integration at the same frequency as they would be during readout. The goal is to mimic the heat generated by the parallel registers during readout in a way that does not corrupt the integrity of the underlying spectra. This approach is very similar to clocked anti-blooming, a technique for reducing the impact of bleeding charge when imaging objects spanning a large range of brightnesses \cite{Janesick1992, Murray2013}. This clocking approach is also very similar to the techniques that have been used to effectively reduce the impact of surface-generated dark current in CCDs for space- and ground-based applications\cite{Burke1991,Jorden2002,Vandersteen2010}. As shown in the clocking diagram in Figure \ref{fig:dither_diagram} and the potential well schematic in Figure 5, during integration phase 4 is always held low to act as a barrier and phase 2 is always held high to collect charge, as during standard operation. In dithered operation the voltages on phases 1 and 3 are varied during integration. With our deep depletion CCD the ``low" voltage is 0 V (the substrate is held at 0 V) and ``high" voltage is +11 V. The amplitude of the dither on phases 1 and 3 can be set so that the full well of the pixel is unchanged, swinging either 11 V or 5.5 V. Based on the requested exposure time, our Archon script executes an integer number of these dithers, with the dither frequency set to match the line time during normal readout (3~ms for our 250 kHz readout mode). The rise time of the transition between high and low states for phases 1 and 3 can also be set so as to minimize the possibility of charge pumping (see Section 5).  We find that the dither clocking has a significant impact on the temperature variations of the CCD package during readout. In Figure \ref{fig:temps} we plot the CCD package temperature during a 500~s integration in both standard and dither modes. A 10~mK transient during integration is reduced to a 1~mK transient. In Figure \ref{fig:seven_days} we show the CCD package temperature in NEID over seven days of gathering calibration spectra. These calibration data included long and short sequences of integrations of wavelength calibration sources. The RMS of the data stream averaged on a 60~s timescale is 0.25~mK.

\section{Potential Pitfalls}

While the dither clocking mode appears very effective at reducing the amplitude of the thermal variations of the CCD package during integration and readout, it is less obvious how this readout mode could impact the RV measurements we are making. We explored several different effects related to the dither clocking that could potentially degrade the quality of the underlying data. We found no evidence for the dither clocking scheme impacting linearity, read noise, or pixel full-well, and dark current remains below 3 e- pix$^{-1}$ hour$^{-1}$ in both modes at our nominal CCD operating temperature of 173~K. 

The movement of charge within a pixel during integration could produce Clock Induced Charge (CIC). This effect is typically more of an issue at much higher clocking frequencies, such as with EMCCDs that are read out very rapidly\cite{Janesick2001}. By comparing a number of long (900~s) dark frames in both standard and dither modes, we found no evidence of CIC at the limit of the dark current and read noise of our system. Given the frequency of the dither clocking, the number of shuffles during these 900~s integrations, and the transition times between the high and low voltage states on the dithered phases, we do not expect to see CIC.

It is possible that the dither clocking scheme could result in a significant signal due to pocket pumping. Typically, pocket pumping is used as a technique to identify traps in the silicon lattice of the CCD. The CCD is exposed to uniform light, then charge is repeatedly shuffled back and forth between adjacent pixels with the illumination turned off. In a perfect CCD, all of the accumulated charge ends up in the pixel it started in, but traps may cause charge to be removed from one pixel and accumulated in another over a large number of transfers. This results in a characteristic dipole pattern that can be used to identify the locations of charge traps\cite{Mostek2010}. Here, a potential concern is a related effect where photoelectrons that are generated under phase 3 when it is low (phase 4 is always low) see an effective electric field that has a very shallow gradient toward phase 2. The result is that a small fraction of these photoelectrons may end up under phase 1 of the adjacent pixel, never to return. This effect can be mitigated by tailoring the clocking profile of phase 3 during shuffle to reduce the total amount of time that photoelectrons see the very flat electric field in the vicinity of phases 3 and 4. We also note that this effect is symmetric between adjacent pixels in that it could occur around phases 3 and 4, but also around phases 1 and 4. In the case where a charge trap is localized to phase 3, this effect could result in a redistribution of charge between pixels that is not, on average, symmetric. We carried out preliminary investigations of this effect with dither clocking using the standard pocket pumping technique of illuminating the device uniformly to approximately half full well and then continuing to integrate in dither mode for 900~s with the light source off. We found no evidence for pocket pumping dipoles in these data. However, this effect and its potential impact on RV measurements must be investigated further. 

The dither clocking mode could result in increased correlations between pixels, due to charge sharing or increased diffusion between pixels resulting from the decreased average barrier voltages. These effects could reduce the Modulation Transfer Function (MTF) of the detector system. We measured MTF in our test Dewar using test pattern projection techniques and laser speckles\cite{laserspeckle}, but found that our optical setup did not produce sufficiently high contrast at high spatial frequencies to measure a difference between the standard and dither modes. Photon Transfer Curves (PTCs) also encode information about correlations between adjacent pixels through the non-linearity of the relationship between variance and flux level\cite{pixelcorrelations}. In Figure 8 we compare measured PTCs in standard and dither modes. We found no significant differences between the PTC properties in the two readout modes. 

It is also possible that the dither clocking alters the physical locations of pixels on the chip as defined by the average gate voltages. While this is not likely an issue if the dither clocking mode is always used when collecting science and calibration data, effects of this type would make it difficult to compare data taken in different modes. We investigated this effect with the CCD operating inside the NEID cryostat by gathering consecutive sets of spectra of flat lamp in standard and dither mode and measuring the average positions of a single spectral order in the cross dispersion (parallel) direction. As shown in Figure 9, we observed shifts of approximately 0.002 pixels between the two modes. While this effect seems robustly measured, it is still to be determined if it is a fixed effect or one that varies over time. If the latter is true, this would be a major source of concern for using the dither clocking for precise RV measurements.  

\section{Conclusions}
Variable heat output from the detector can be a major source of systematic measurement error for precise RV spectrometers. Radiative coupling between the CCD and the instrument optics, as well as physical deformations of the CCD, can lead to instrumental shifts that are large compared to the RV signals from Earth-like planets orbiting sun-like stars. Given that CCDs in modern high-resolution spectrometers are large, up to nearly 100 cm$^2$, and the variations in heat loads between idle, integration, and readout may be hundreds of mW, actively controlling the temperature of the CCD becomes important. We have developed a technique to reduce the overall variations of the CCD package temperature in the NEID spectrometer by dithering the parallel and serial clocks during integration. We find that this clocking scheme reduces the overall variation of the CCD package temperature to the mK level compared to thermal transients at the 20~mK level in standard operating mode. We have explored several different ways in which this clocking scheme could corrupt the underlying spectral data. Initial investigations revealed no significant effects, but ultimately the stability of the instrument as measured through observations of the laser frequency comb will be necessary to evaluate the impact of the dither clocking on RV performance.  

\acknowledgments 
We thank the organizers of ISPA-2018 for organizing a very informative workshop. This work was carried out in part at the Singh Center for Nanotechnology at the University of Pennsylvania, which is supported by the NSF National Nanotechnology Coordinated Infrastructure Program under grant NNCI-1542153. NEID is funded by NASA through JPL by contract 1547612. Mark Giovinazzi is supported by an NSF Graduate Research Fellowship. The authors would like to thank an anonymous referee for insightful suggestions that helped to improve this manuscript.

\section*{Disclosures}
The authors have no conflicts of interest to declare. 

\clearpage 

\clearpage

\begin{figure}
\begin{center}
\begin{tabular}{c}
\includegraphics[height=12cm]{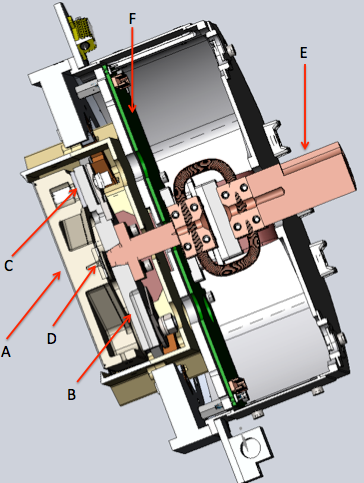}
\end{tabular}
\end{center}
\caption 
{ \label{fig:xsection}
Cross section of the NEID CCD mount. (A) CCD face, (B) Mo cold block, which has a single PT100 RTD embedded, (C) Thermal contact between CCD and Mo cold block is through three polished Invar pads on the SiC package, (D) A PT100 RTD in a small copper block is attached to the center of the SiC package using a single M4 screw, (E) A portion of the Cu cold linkage attaching the Mo block to the instrument LN2 reservoir, and (F) CCD electronics board.} 
\end{figure} 

\clearpage

\begin{figure}
\begin{center}
\begin{tabular}{c}
\includegraphics[width=16cm]{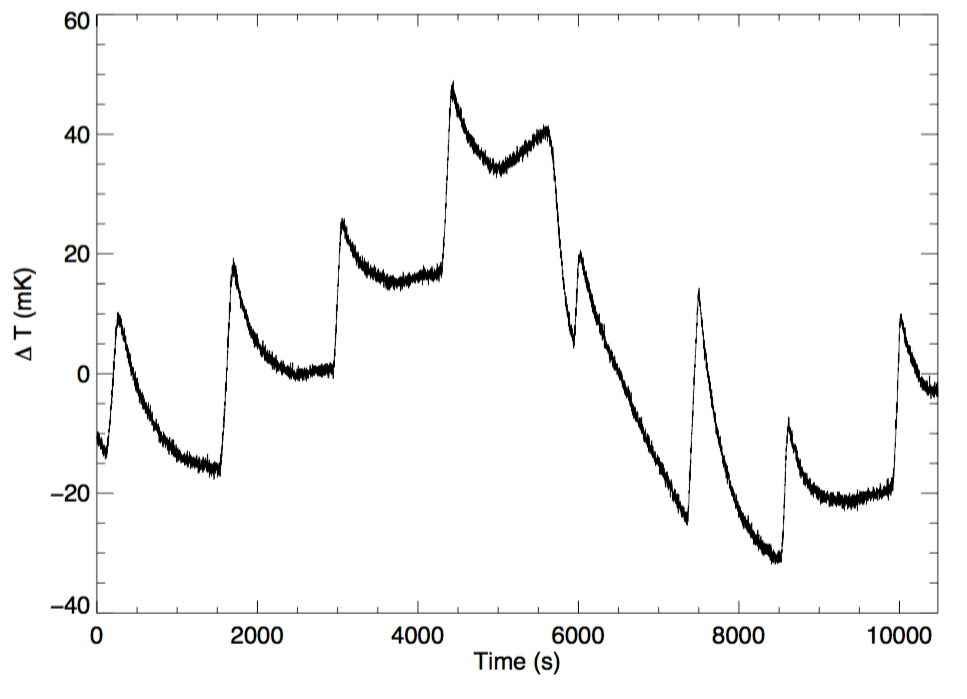}
\end{tabular}
\end{center}
\caption 
{ \label{fig:long_temp}
Changes in the CCD package temperature in the test Dewar during a series of calibration exposures of different lengths gathered over 2.75 hours in standard clocking mode. Here, both the parallel and serial registers are cycled during idle, so a decrease in package temperature corresponds to the start of an integration and a rapid rise in package temperature corresponds to the beginning of readout. We observed long-term trends in CCD temperature due to variations in the test Dewar fill level, but the Dewar hold time is long compared to the 2.75 hour data stream shown here.} 
\end{figure} 

\clearpage 

\begin{figure}
\begin{center}
\begin{tabular}{c}
\includegraphics[width=14cm]{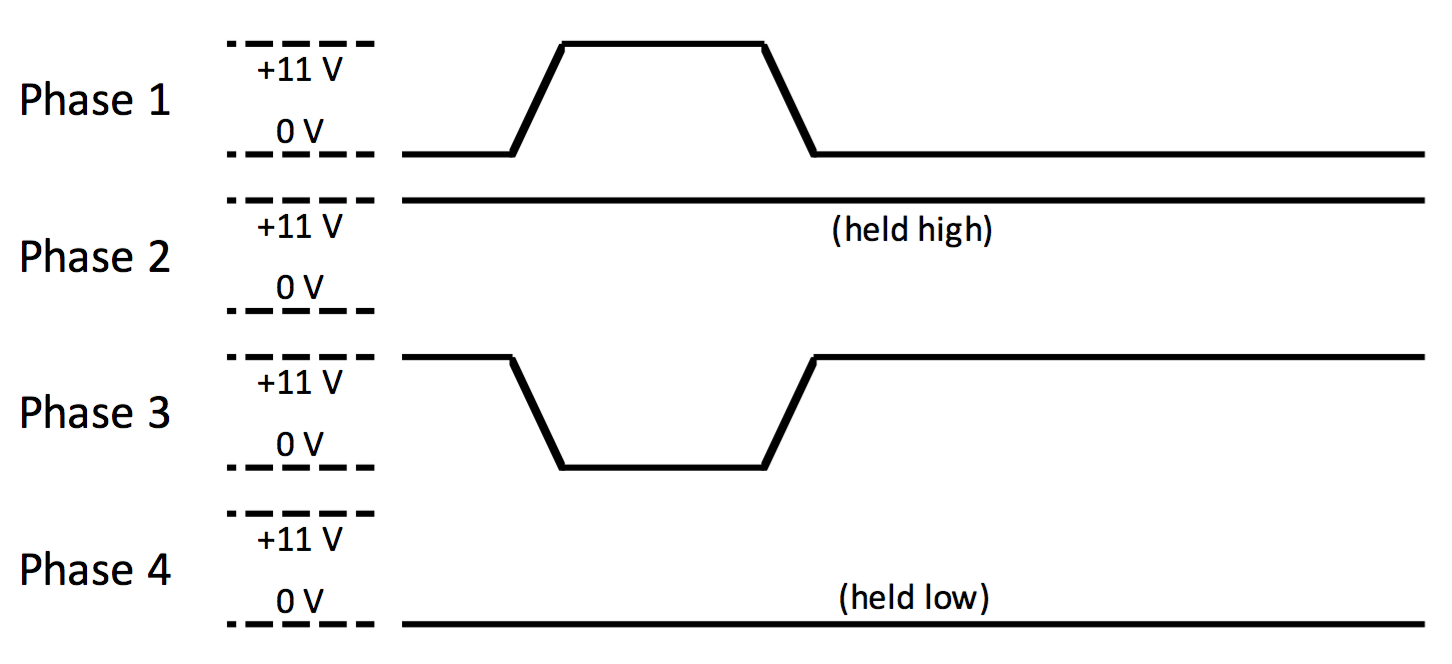}
\end{tabular}
\end{center}
\caption 
{ \label{fig:dither_diagram}
Schematic showing the dither clocking scheme. During non-dither operation, phases 1 and 4 would be held at 0~V and phases 2 and 3 would be held at +11 V during the entire integration.  } 
\end{figure} 

\clearpage

\begin{figure}
\begin{center}
\begin{tabular}{c}
\includegraphics[width=12cm]{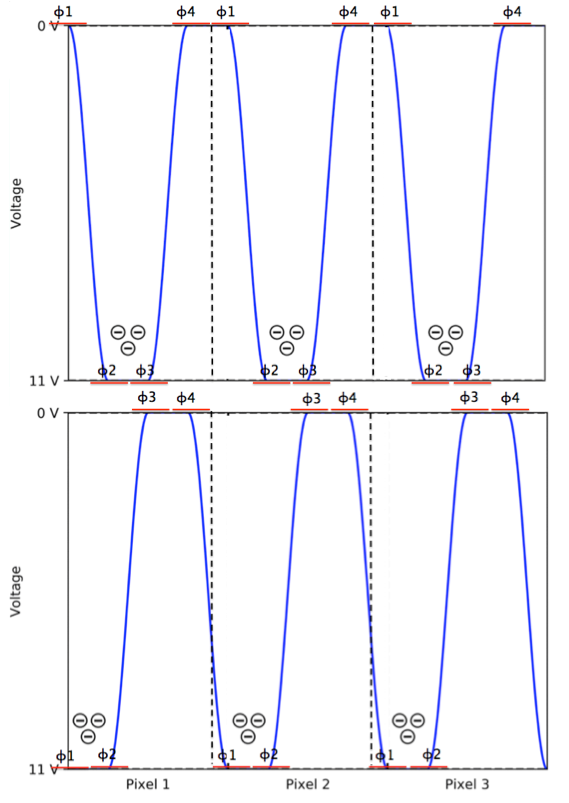}
\end{tabular}
\end{center}
\caption 
{ \label{fig:potentials}
Schematic of change in potentials defining CCD pixels at the two extremes of the dither process, where phases 1 and 3 swap from 0~V to 11~V. The blue lines represent the potential wells seen by the photoelectrons and the dashed vertical lines represent the corresponding pixel boundaries.} 
\end{figure} 

\clearpage 

\begin{figure}
\begin{center}
\begin{tabular}{c}
\includegraphics[width=13cm]{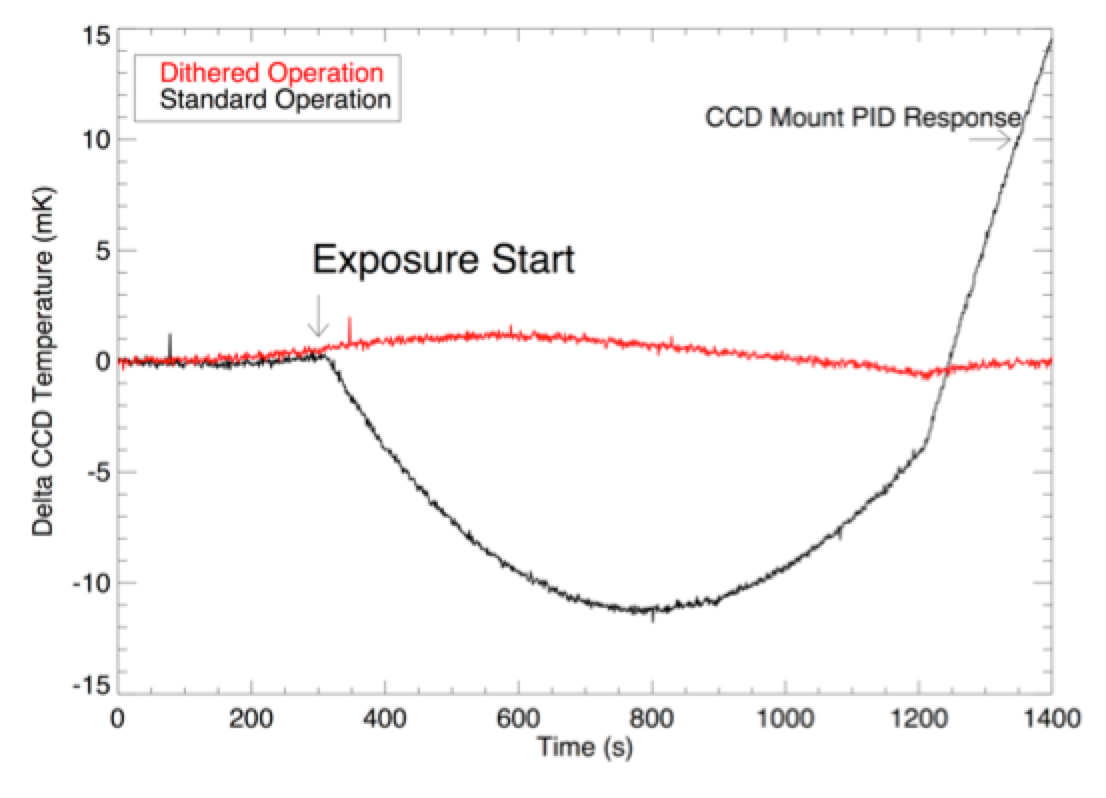}
\end{tabular}
\end{center}
\caption 
{ \label{fig:temps}
Comparison of CCD package temperature changes during a 500~s exposure in dither and standard modes in the test Dewar.} 
\end{figure} 

\clearpage

\begin{figure}
\begin{center}
\begin{tabular}{c}
\includegraphics[width=16cm]{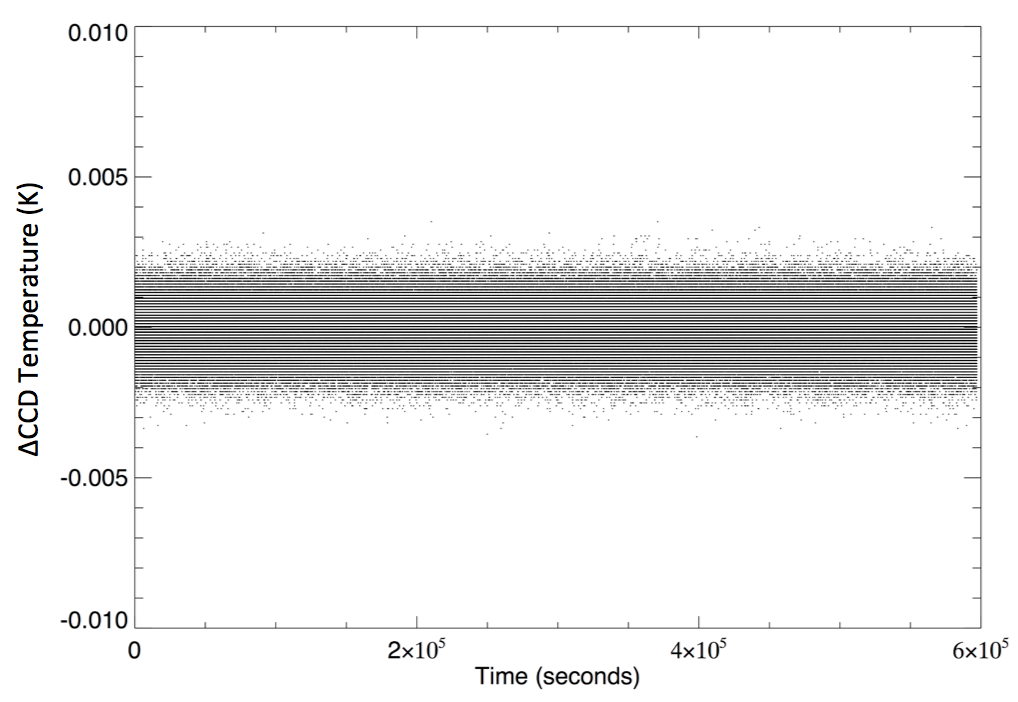}
\end{tabular}
\end{center}
\caption 
{ \label{fig:seven_days}
Change in temperature of the CCD package over seven days of operation in NEID. During the period, NEID was gathering a wide range of different calibration exposures in the laboratory, with a duty cycle similar to what may be experienced during operations in the field. The overall RMS of the data stream after binning on a 60~s timescale is 0.25~mK.} 
\end{figure} 

\clearpage

\begin{figure}
\begin{center}
\begin{tabular}{c}
\includegraphics[width=14cm]{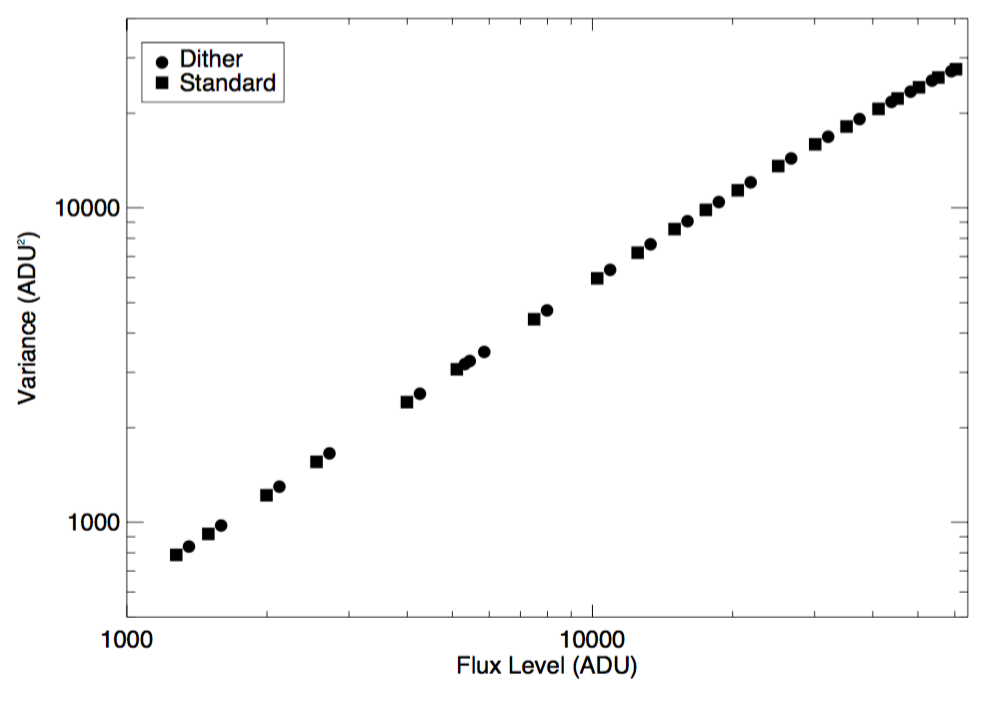}
\end{tabular}
\end{center}
\caption 
{ \label{fig:ptc}
Comparison of Photon Transfer Curves (PTCs) generated in dither and standard modes. The differences between the two best-fit third-order polynomials to these PTCs are less than 0.2$\%$ over this flux range. } 
\end{figure} 

\clearpage

\begin{figure}
\begin{center}
\begin{tabular}{c}
\includegraphics[width=14cm]{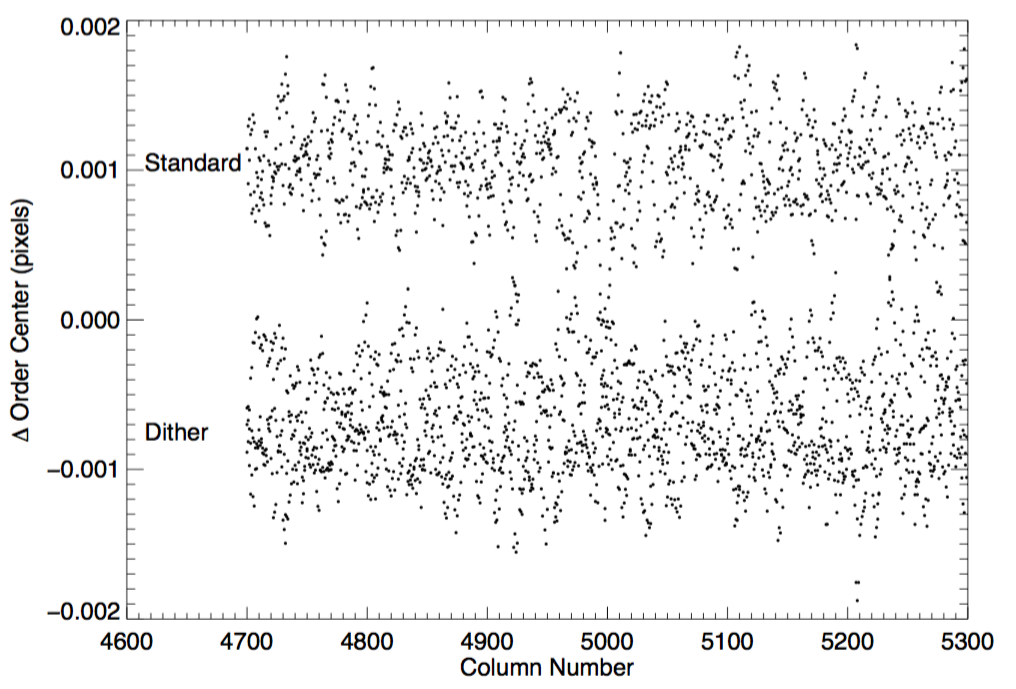}
\end{tabular}
\end{center}
\caption 
{ \label{fig:shifts}
Measured position of a single spectral order over a series of exposures in the dither and standard modes in NEID.} 
\end{figure} 

\clearpage 


\bibliography{report}   
\bibliographystyle{spiejour}   



\end{spacing}
\end{document}